\documentclass[reprint, twocolumn, amsfonts, amsmath, amssymb, superscriptaddress, prl]{revtex4-2}

\usepackage{graphicx}
\usepackage{dcolumn}
\usepackage{bm}
\usepackage[normalem]{ulem}
\usepackage{mathtools}
\usepackage[dvipsnames]{xcolor}
\usepackage[linkcolor=Blue,citecolor=Blue,urlcolor=Blue,colorlinks=true]{hyperref}

\renewcommand{\vec}{\boldsymbol}
\newcommand{\mat}{\boldsymbol}

\newcommand*{\fullref}[1]{\hyperref[{#1}]{\ref*{#1} ``\nameref*{#1}''}}

\begin{document}

\title{Delayed excitations induce polymer looping and coherent motion}

\author{Andriy Goychuk}
\email{andriy@goychuk.me}
\affiliation{Institute for Medical Engineering and Science, Massachusetts Institute of Technology, Cambridge, MA 02139, United States}%
\author{Deepti Kannan}
\affiliation{Department of Physics, Massachusetts Institute of Technology, Cambridge, MA 02139, United States}%
\author{Mehran Kardar}
\email{kardar@mit.edu}
\affiliation{Department of Physics, Massachusetts Institute of Technology, Cambridge, MA 02139, United States}%

\date{\today}

\begin{abstract}
We consider inhomogeneous polymers driven by energy-consuming active processes which encode temporal patterns of athermal kicks.
We find that such temporal excitation programs, propagated by tension along the polymer, can effectively couple distinct polymer loci.
Consequently, distant loci exhibit correlated motions that fold the polymer into specific conformations, as set by the local actions of the active processes and their distribution along the polymer.
Interestingly, active kicks that are canceled out by a time-delayed echo can induce strong compaction of the active polymer.
\end{abstract}

\maketitle


The biological function of a polymer is tightly connected to its conformational statistics.
Longstanding research has illuminated the mechanisms of sequence-controlled polymer folding near thermal equilibrium~\cite{Review::Onuchic1997, Review::Pande2000, Review::Shakhnovich2006, Review::Dill2008, Review::Wolynes2015, Review::Nassar2021}.
In such passive systems, the entropic price for folding is paid by monomer-solvent interactions~\cite{Article::England2008, England2011AllosteryEffects, Article::Perunov2014} and by monomer-monomer interactions, which can encode complex potential landscapes~\cite{Review::Onuchic1997, Review::Wolynes2015}.
Similar ideas have been successfully applied to the folding of chromatin, a large heteropolymer consisting of genomic DNA and associated proteins~\cite{Review::Misteli2020}.
However, chromatin also shows features of active systems, where nonequilibrium fluctuations~\cite{Article::Weber2012} and coherent motion~\cite{Article::Zidovska2013, Article::Backlund2014, Article::Shaban2018} emerge only in presence of adenosine triphosphate (ATP).

ATP fuels molecular motor activity and biochemical reactions among chromatin-associated proteins~\cite{Review::Narlikar2013, Review::Uhlmann2016, Review::Reid2017}.
For example, SMC complexes~\cite{Review::Uhlmann2016} share features with other active biochemical systems~\cite{Review::Halatek2018} by cycling between an ADP-bound and an ATP-bound state.
Such rapid switching in conformation, interactome and biochemical affinity of chromatin-associated proteins suggests that chromatin potential landscapes actively fluctuate.
On a coarse-grained level, these chemically fueled fluctuations can be viewed as local excitations which kick the polymer out of thermal equilibrium.
The active kicks at different loci will vary in magnitude and could even show sequence-specific correlations, if driven by long-ranged bulk protein concentration or electrostatic potential gradients~\cite{Article::Goychuk2023}.

Active processes can elicit polymer folding~\cite{Article::Goychuk2023}, but cannot be discerned from passive coupling of distant loci via additional Hookean springs~\cite{Article::LeTreut2018, Article::Shi2019} based on contact frequency data alone.
This structural equivalency allows quasi-equilibrium theories to absorb energy-consuming processes into effective interaction parameters~\cite{Perspective::Zhang2017, Article::DiPierro2018}.
However, active and passive dynamics are expected to significantly differ~\cite{Review::Brangwynne2009, Article::Gnesotto2018} in response to ATP~\cite{Article::Weber2012, Article::Zidovska2013, Article::Backlund2014, Article::Shaban2018}.
To that end, mechanical stress propagation is crucial, because it coordinates the motion of chromosomal loci either via the backbone~\cite{Article::Lampo2016, Article::Put2019}, by chemical interactions among non-neighboring monomers~\cite{Article::DiPierro2018, Article::Salari2022, Article::Liu2018, Article::Jiang2022}, or via the surrounding fluid~\cite{Article::Bruinsma2014, Article::Eshghi2023}.
This suggests local active kicks, for example due to motor activity~\cite{Article::Put2019, Article::Bruinsma2014, Article::Saintillan2018, Article::Mahajan2022}, or persistent active motion of specific loci~\cite{Article::Ghosh2014, Kaiser2015, Article::Eisenstecken2016, Article::Eisenstecken2017a, Article::Mousavi2019, Article::Anand2020, Article::Brahmachari2023, Article::Eisenstecken2017b, Article::Osmanovic2017, Osmanovic2018, Article::Liu2018, Article::Saito2019, Article::Jiang2022, Article::Ghosh2022, Article::Ghosh2022b, Article::Brahmachari2023, Review::Winkler2020, Article::Dutta2024} such as biomolecular condensates~\cite{Hanczyc2011, Mirco2023, JambonPuillet2023, Article::Schede2023, Article::Demarchi2023}, to have non-local consequences.
In contrast to previously studied systems with exponential memory, however, we hypothesize that active force fluctuations, due to binding and unbinding of individual molecular motors~\cite{Article::MacKintosh2008} or chemical reaction cycles, have specific temporal signatures.
Hence, in this letter, we show how an interplay between sequence-specific temporal patterns of excitations and stress propagation can induce correlated motion and folding.

Towards a theoretical framework for such inquiry, we idealize the polymer as a space curve $\vec{r}(s,t)$ parametrized by dimensionless material coordinates ${s \in [-L/2,L/2]}$.
Tension propagates along the polymer via lateral stretching modes or via transversal bending modes, as captured by the energetics of an inhomogeneous wormlike chain with Kuhn length ${b}$. 
Assuming dissipative relaxation of the polymer over a microscopic monomer diffusion time $\tau_b$, and active kicks captured through a random velocity field $\vec{\eta}(s,t)$, the dynamics are described by
\begin{equation}
     \tau_b \bigl[\partial_t\vec{r}(s,t) {-} \vec{\eta}(s,t)\bigr] {\,=\,} \partial_s \bigl[ \kappa(s) \partial_s\vec{r}(s,t) \bigr] {-} \tfrac{1}{4} \partial_s^4\vec{r}(s,t) .
\label{eq:langevin_real_wlc}
\end{equation}
In the limit of uniform activity and ${\kappa(s)=\kappa=1}$, the above wormlike chain has steady-state tangent-tangent correlations decaying as $\langle \vec{\tau}(s,t)\cdot \vec{\tau}(s',t)\rangle=e^{-|s-s'|/l_p}$, with ${\vec{\tau}(s,t) \coloneqq b^{-1} \partial_s \vec{r}(s,t)}$,
and dimensionless persistence $l_p=1/\sqrt{4\kappa}$ along the sequence.
In the presence of nonuniform activity, we employ a spatially varying $\kappa(s)$ to enforce an ensemble-averaged and hence weak constraint of local inextensibility, ${\langle |\vec{\tau}(s,t)|^2 \rangle = 1}$. 
Such weak length constraints are certainly different from strongly enforcing inextensibility on the level of individual trajectories, for example, through elastic constraints, which are known to qualitatively affect tension propagation~\cite{Article::Hallatschek2007}.
These limitations notwithstanding, we expect that our simplified theory can provide insights into the folding and dynamics of polymers driven by temporal patterns of excitations.
In the following, for convenience and clarity of these ideas, the main text presents analytic theory in the limit of continuous polymers, while figures show numeric evaluations of the analytic theory for discrete chains as described in the Supplemental Material (SM)~\cite{supplement}.

\begin{figure}[t]
    \centering
	\includegraphics{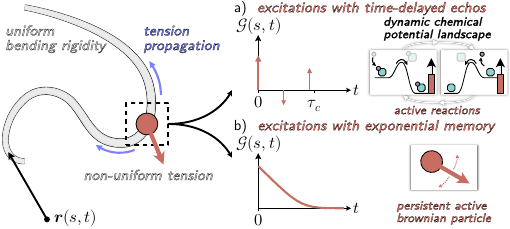}
	\caption{%
    Sketch of an active semiflexible polymer.
    Active processes generate excitations, which propagate along the polymer backbone.
    A non-uniform profile of activity leads to non-uniform line tension, which maintains chain inextensibility.
    In the present work, we consider two types of excitations: 
    a) Time-delayed echos of past kicks at the same location, such as binding followed by unbinding events.
    b) Exponentially correlated kicks by persistent self-propelling particles.
	}
	\label{fig::model}
\end{figure}

To describe temporal patterns of active kicks, we introduce a scalar memory kernel $\mathcal{G}(s, t)$ with characteristic time $\tau_c$, and construct non-Markovian excitations
%
\begin{equation}
\vec{\eta}(s,t) = 
\sqrt{\frac{2b^2}{\tau_{b}}}
\lim_{\varepsilon \to 0} 
\int_{-\infty}^{t+\varepsilon} \!\! d\tau \!
\begin{bmatrix}
\mathcal{G}(s, t-\tau) \\
\delta(t-\tau)
\end{bmatrix} 
\!\cdot\!
\begin{bmatrix}
  \vec{\gamma}^{a}(s,\tau) \\
  \vec{\gamma}^{p}(s,\tau)
\end{bmatrix} 
\!,
\label{eq:memory}
\end{equation}
%
%
whose active (a), and passive (p), contributions are seeded by Gaussian noises $\vec{\gamma}^{a/p}(s,t)$ with zero mean.
In the simplest scenario, both contributions have the same characteristic length scale, $\langle\vec{\gamma}^\alpha(s,t) \cdot \vec{\gamma}^\beta(s',t') \rangle = \delta_{\alpha\beta}/(2l_e) \, e^{-|s-s'|/l_e} \, \delta(t-t')$.
The ratio $l_e/l_p$ between the correlation length of the excitations and the polymer is, as explained in this article, a key parameter.
In the limit $l_e \rightarrow 0$,  excitations are independent along the sequence, while a finite $l_e$ indicates that nearby loci are typically kicked in similar directions.
We only consider cases where the variations in $\mathcal{G}(s,t)$ are smooth over the correlation length of the excitations, ${\partial_s\mathcal{G} \ll l_e^{-1} \mathcal{G}}$, so the covariance is well approximated by
\begin{subequations}
\label{eq:covariance_excitations}
\begin{equation}
    \langle\vec{\eta}(s, t) \cdot \vec{\eta}(s', t')\rangle = \frac{2b^2}{\tau_b} 
    \frac{1}{2l_e} \, e^{-\frac{|s-s'|}{l_e}} 
    \widehat{\mathcal{C}}\bigl(\tfrac{s+s'}{2}, t-t'\bigr) \, ,
\end{equation}
with
\begin{equation}
    \widehat{\mathcal{C}}(s,t) = \delta(t) + \lim_{\varepsilon \to 0}  \int_{-\varepsilon}^{\infty} \! d\tau \, \mathcal{G}(s, |t| + \tau) \, \mathcal{G}(s, \tau) \, .
\end{equation}
\end{subequations}
This quantifies how the covariance of athermal excitations~\cite{Article::Osmanovic2017} explicitly depends on their temporal patterns.
We consider two choices of $\mathcal{G}(s,t)$, one which describes heteropolymers assembled from persistent random walkers~\cite{Article::Ghosh2014, Kaiser2015, Article::Eisenstecken2016, Article::Eisenstecken2017a, Article::Eisenstecken2017b, Article::Osmanovic2017, Osmanovic2018, Article::Mousavi2019, Article::Saito2019, Article::Anand2020, Article::Ghosh2022, Article::Ghosh2022b, Article::Brahmachari2023, Review::Winkler2020}, and one where polymers are subject to timed sequences (temporal program) of excitations [Fig.~\ref{fig::model}].

Having specified the nature of the excitations, we next analyze Eq.~\eqref{eq:langevin_real_wlc} via a Rouse mode decomposition~\cite{Article::Rouse1953}, where ${\vec{r}(s, t) \rightarrow \tilde{\vec{r}}_q(t)}$ and ${\vec{\eta}(s, t) \rightarrow \tilde{\vec{\eta}}_q(t)}$ indicate Fourier transforms.
We compactify the notation by concatenating all Rouse modes row-wise into a matrix $\mat{R}(t)$ with rows ${R_{q,\dots}(t) \coloneqq \tilde{\vec{r}}_q(t)}$, and analogously define the random velocity mode matrix $\mat{H}(t)$ with rows ${H_{q,\dots}(t) \coloneqq \tilde{\vec{\eta}}_q(t)}$.
The Fourier-transformed Eq.~\eqref{eq:langevin_real_wlc} is solved by
\begin{equation}
    \mat{R}(t) = \int_{-\infty}^{t} \! d\tau \, e^{-\mat{J} (t-\tau)} \cdot \mat{H}(\tau) \, ,
    \label{eq::solution_formal}
\end{equation}
with the response matrix, $\mat{J}$, given by $\tau_b J_{qk} = \frac{1}{L} q k \, \kappa_{q-k} + \tfrac{1}{4} q^4 \delta_{qk}$.
In principle, one could also consider an additional memory kernel that acts on the backbone velocities, $\partial_t \vec{r}(s,t)$, and approximates the viscoelasticity of the surrounding medium~\cite{Article::Weber2010, Article::Weber2010b, Article::Lampo2016}.
The interplay between the timescales of the system could then lead to further interesting effects. 
Even in the simpler scenario, however, there is already an interplay between the memory of the excitations and the relaxation time of the polymer.

First, we analyze the dynamics of a polymer  subjected to generic excitations with covariance ${\mat{C}(t-t') \coloneqq \langle\mat{H}(t)\cdot\mat{H}^\dagger(t')\rangle}$ and characteristic decorrelation time $\tau_c$, so that ${\mat{C}(\tau) \sim 0}$ for $|\tau| > \tau_c$.
In the limit of either long timescales, ${|t-t'| > \tau_c}$, or in steady state, ${t = t'\to\infty}$, the memory of the excitations can be effectively integrated out~\cite{supplement}, because the polymer has ample time to run through many iterations of its excitation program $\mathcal{G}$.
More specifically, one finds that the polymer dynamics resemble that of a chain where the temporal excitation programs are replaced by couplings between different excitation modes, ${\langle \mat{H}(t) \cdot \mat{H}^\dagger(t') \rangle \approx \mat{C}_\textit{eff} \, \delta(t-t')}$ with
\begin{equation}    
    \mat{C}_\textit{eff} \coloneqq \int_{0}^{\infty} \!\!\! d\tau \, \left[ \mat{C}(\tau) \cdot e^{-\mat{J}^\dagger \tau} 
    + e^{-\mat{J} \tau} \cdot \mat{C}^\dagger(\tau) \right]     
    \, .
\label{eq::correlation_effective_spectral}
\end{equation}
The covariance $\mat{X}(0)\equiv \langle \mat{R}(t) \cdot \mat{R}^\dagger(t) \rangle$ of the steady-state distribution of conformations is then implicitly defined by the Lyapunov equation ${\mat{J}\cdot \mat{X}(0) + \mat{X}(0) \cdot \mat{J}^\dagger = \mat{C}_\textit{eff}}$.
In the limit that the memory is short-lived, ${\tau_c \ll \tau_b}$, the polymer fluctuations are well approximated by ${\langle \mat{R}(t) \cdot \mat{R}^\dagger(t') \rangle \approx  e^{-\mat{J} \, |t-t'|} \cdot \mat{X}(0)}$ for ${t > t'}$.
The complementary case where the memory is long-lived, ${\tau_c \gtrsim \tau_b}$, however, can be only obtained by evaluating the analytical solutions numerically.
We next analyze specific models for the response matrix of the polymer, $\mat{J}$, and for the temporal pattern of excitations encoded in $\mat{C}(\tau)$.

\begin{figure}[t]
    \centering
	\includegraphics{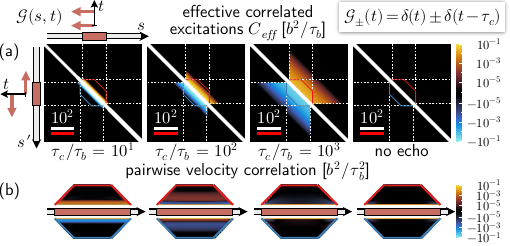}
	\caption{%
    (a) Local active processes lead to effectively correlated excitations of a discrete $10^3$-mer flexible chain.
    Red arrows indicate that each active kick in the active region ($\epsilon = 1$ for $s \in$ red segment, delimited by dashed lines) is followed by a time-lagged positive ($\mathcal{G}_+$, above diagonal) or negative ($\mathcal{G}_-$, below diagonal) echo.
    Scale bars: $10^2$ monomers (white) or Kuhn segments (red).
    (b) Same-time velocity pair correlations, $\langle \vec{v}^\delta(s,t) \cdot \vec{v}^\delta(s',t) \rangle$ with time discretization $\delta = 0.1\tau_b \ll \tau_c$ shorter than the memory.
    Red and blue outlines correspond to solid lines in (a).
    Positive (negative) echos with larger lag time $\tau_c$ cause longer-ranged but weaker positive (negative) velocity correlations.
    }
    \label{fig::time-delayed_Rouse}
\end{figure}

In the simplest scenario, on length scales much larger than the noise correlation length and the persistence length, ${\Delta s \gg \max(l_e,2l_p)}$, tension is expected to propagate predominantly through longitudinal modes.
In this limit, it is tempting to approximate $\langle\vec{\gamma}^\alpha(s,t) \cdot \vec{\gamma}^\beta(s',t') \rangle \approx \delta_{\alpha\beta} \, \delta(s-s') \, \delta(t-t')$, set $\kappa(s) = 1$, and drop the bending term in Eq.~\eqref{eq:langevin_real_wlc}, leading to $\tau_b J_{qk} \sim q^2 \delta_{qk}$. 
Transforming Eq.~\eqref{eq::correlation_effective_spectral} from Fourier space to sequence space, one finds 
\begin{equation}
    \mathcal{C}_\textit{eff}(s,s') = \frac{2b^2}{\tau_b} \int_{0}^{\infty}\! d\tau \, \frac{\widehat{\mathcal{C}}(s,\tau) + \widehat{\mathcal{C}}(s',\tau)}{\sqrt{4\pi \tau/\tau_b}} \, e^{- \frac{(s-s')^2}{4\tau/\tau_b}} \, ,
\label{eq:effective_excitations}
\end{equation}
which, as will be illuminated shortly, is a good approximation when ${l_e \gg l_p}$.
Thus, purely local temporal sequences of kicks, which are relayed via the line tension of the polymer backbone, can lead to effective sequence-controlled correlated excitations.

This is illustrated in Fig.~\ref{fig::time-delayed_Rouse}(a) for a discrete flexible chain subjected to inhomogeneous excitations $\mathcal{G}(s,t) = \sqrt{\epsilon(s)} \, \mathcal{G}_\pm(t)$ with echos, $\mathcal{G}_\pm(t) = \delta(t) \pm \delta(t - \tau_c)$.
Within the lag time $\tau_c$ between an active kick and its subsequent echo, forces propagate diffusively across a typical distance $|s-s'| \propto \sqrt{\tau_c / \tau_b}$ [Eq.~\eqref{eq:effective_excitations}], effectively coupling the excitations at loci $s$ and $s'$.
Consequently, the locus velocities $\vec{v}^\delta(s,t) \coloneqq [\vec{r}(s,t+\delta) - \vec{r}(s,t)] / \delta$ also exhibit pairwise correlations which are well approximated by $\langle \vec{v}^\delta(s,t) \cdot \vec{v}^\delta(s',t) \rangle \approx \delta^{-1} \mathcal{C}_\textit{eff}(s,s')$ for $\tau_c \ll \delta \ll \tau_b$.
This approximation fails if the time discretization $\delta$ is too fine to integrate out the memory, i.e. when $\delta \ll \tau_c$.
Nevertheless, when compared to an explicit evaluation of the pairwise velocity correlations, $\mathcal{C}_\textit{eff}(s,s')$ reasonably predicts which regions show coordinated motion [left two panels in Fig.~\ref{fig::time-delayed_Rouse}(b)].
As discussed in our prior work~\cite{Article::Goychuk2023}, because correlated (anticorrelated) excitations imply effective attraction (repulsion) between loci, Eq.~\eqref{eq:effective_excitations} also informs about the conformations of the polymer.

To further elucidate the results so far, we  now revisit a scenario where the excitations have exponential memory, $\mathcal{G}(s,t) = \sqrt{\epsilon(s)} \, \tau_c^{-1} e^{-|t|/\tau_c}$, which has been the focus of several theoretical studies~\cite{Article::Ghosh2014, Kaiser2015, Article::Eisenstecken2016, Article::Eisenstecken2017a, Article::Eisenstecken2017b, Article::Osmanovic2017, Osmanovic2018, Article::Mousavi2019, Article::Saito2019, Article::Anand2020, Article::Ghosh2022, Article::Ghosh2022b, Article::Brahmachari2023} and reviewed in Ref.~\cite{Review::Winkler2020}.
Diffusive propagation of tension through lateral stretching modes [Eq.~\eqref{eq:effective_excitations}] leads to 
\begin{equation}
    \mathcal{C}_\textit{eff}(s,s') = \frac{2b^2}{\tau_b} \left[
    \delta(s-s') 
    + \frac{\epsilon(s)+\epsilon(s')}{4 l_\epsilon} \, e^{-\frac{|s-s'|}{l_{\epsilon}}}
    \right] \, ,
\end{equation}
with $l_\epsilon \coloneqq \sqrt{\tau_c / \tau_b}$.
For a flexible chain, a local increase in activity should then lead to swelling~\cite{Article::Goychuk2023, Review::Winkler2020}.
In contrast, semiflexible active chains can show either contraction or swelling when the average activity is increased~\cite{Review::Winkler2020}.
To reconcile these results, we next study the propagation of transversal excitation modes via bending, which we have neglected so far.

\begin{figure}[t]
    \centering
	\includegraphics{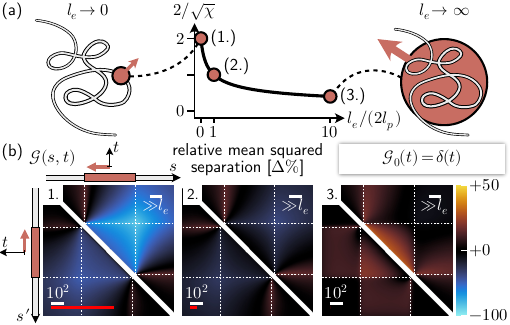}
	\caption{%
    Response of a semiflexible chain to active processes depends on the noise correlation length $l_e$.
    (a) By agitating bending modes, non-uniform activity can induce tension modulations [Eq.~\eqref{eq:langevin_real_effective_rouse}], quantified by $2/\sqrt{\chi}$ with control parameter $l_e$.
    The limit $l_e\to\infty$ recovers a flexible chain with uniform tension. 
    (b) Change in pairwise squared separation due to memory-less active kicks in an active region ($\epsilon = 0.5$ for $s \in$ red segment, delimited by dashed lines), compared to a uniform chain.
    \emph{Above diagonal}: Continuous wormlike chain for different $l_e$, on arbitrary length scales $\gg l_e$ (scale bar).
    \emph{Below diagonal}: Discrete semiflexible chain of $10^3$ coarse-grained monomers ($10^{2}$, $10^{3}$, or $10^{6}$ Kuhn lengths) which receive independent kicks.
    Scale bars: $10^2$ monomers (white) or persistence lengths (red).
    (1.) For $l_e \rightarrow 0$, agitation of bending modes effectively \emph{compacts} the polymer.
    (2.) For $l_e = 2l_p$, the active polymer resembles an inextensible \emph{flexible} chain.
    (3.) For $l_e \rightarrow \infty$, the polymer behaves like an extensible flexible chain with agitation of stretching modes.
    }
    \label{fig::wormlike-chain_model-classes}
\end{figure}

Assuming that the activity is large and nearly homogeneous, $\epsilon(s) \gg 1$ and $\partial_s \epsilon(s) \ll l_\epsilon^{-1} \epsilon(s)$, the exponential memory can be replaced by memory-less noise of the form $\mathcal{G}(s,t) = \sqrt{\epsilon(s)} \, \delta(t)$ and a renormalized correlation length $l_e \approx l_\epsilon$ in Eq.~\eqref{eq:covariance_excitations}.
We determine the local Lagrange multiplier $\kappa(s)$ from the weak constraint $\langle |\vec{\tau}(s,t)|^2 \rangle = 1$, as shown in the SM~\cite{supplement}.
On large length scales (i.e., for long wave modes), the dynamics of the polymer are  approximately given by
\begin{equation}
     \tau_b\bigl[\partial_t\vec{r}(s,t) - \vec{\eta}(s,t) \bigr] = \partial_s \! \left[ \left(\frac{1}{\chi} + \frac{2\epsilon(s)}{\chi^{3/2}} \right) \partial_s\vec{r}(s,t) \right] \!,
\label{eq:langevin_real_effective_rouse}
\end{equation}
%
where the adjustment $\chi \equiv \chi(l_e)$ rescales the line tension as derived in the SM~\cite{supplement}.
The reduction in line tension ensures that the polymer does not contract due to the finite correlation length of the excitations~\cite{Article::Goychuk2023}.
Figure~\ref{fig::wormlike-chain_model-classes}(a) shows the relative weight of the inhomogeneous term in the square brackets of Eq.~\eqref{eq:langevin_real_effective_rouse}, which for $l_e \gg l_p$ decays as $1/\sqrt{l_e}$.
Thus, treating the polymer as an effective Rouse chain is internally consistent~\footnote{One then recovers the dynamics of a flexible chain by rescaling $s \mapsto \tilde{s}/\chi$ and $t \mapsto \tilde{t}/\chi$ and dropping the tildes.} if the excitations decorrelate over distances much larger than the Kuhn length [Fig.~\ref{fig::wormlike-chain_model-classes}(b)].
Then, large coiled segments (Pincus blobs~\cite{RubinsteinBook}) experience coherent excitations, primarily leading to agitation of longitudinal modes.
%

%

In general, however, Eq.~\eqref{eq:langevin_real_effective_rouse} demonstrates that regions with locally increased activity also show an increased tension of the polymer backbone.
For cases where the excitations occur on length scales smaller than a Pincus blob, $l_e < 2l_p$, this can qualitatively change the response of the polymer by causing active segments to contract instead of expand [above the diagonal in Fig.~\ref{fig::wormlike-chain_model-classes}(b)].
These predictions of continuum theory also apply to discrete semiflexible polymers [below the diagonal in Fig.~\ref{fig::wormlike-chain_model-classes}(b)], where the response matrix is described in the SM~\cite{supplement} and in Refs.~\cite{Article::Winkler2003, Article::Tejedor2022}.
Mechanistically, these effects can be traced back to the agitation of transversal bending modes, which inevitably lead to contraction when coupled with a length constraint.
Thus, the interplay between features of the athermal excitations and the microscopic mechanics of the polymer can determine if active regions expand ($l_e > 2l_p$) or contract ($l_e < 2l_p$).
Extrapolating these results, the persistence time $\tau_c$ of active particles can control if semiflexible chains swell or contract~\cite{Review::Winkler2020}. 

%
\begin{figure}[t]
    \centering
	\includegraphics{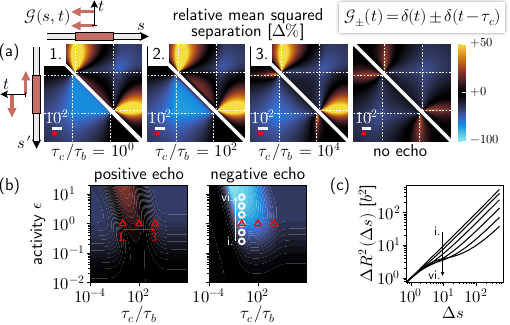}
	\caption{%
    Response of a discrete $1000$-mer semiflexible chain to active processes. 
    (a) Change in pairwise squared separation compared to a uniform chain.
    Red arrows illustrate that each active kick in the active region ($\epsilon = 1$ for $s \in$ red segment, delimited by dashed lines) is followed by a time-lagged positive ($\mathcal{G}_+$, above diagonal) or negative ($\mathcal{G}_-$, below diagonal) echo.
    Scale bar: $10^2$ monomers (white) or persistence lengths (red).
    (b) Average change in pairwise separation within the active segment as a function of the activity $\epsilon$ and the echo lag time.
    Time-delayed echos show strongest effect for intermediate time lags ($\tau_c \sim \tau_b$).
    In the limits $\tau_c \ll \tau_b$ or $\tau_c \gg \tau_b$, excitations become statistically independent.
    Red triangles correspond to mean squared separation maps in (a).
    (c) Mean squared distance from the center monomer in a homogeneous semiflexible chain driven by kicks with echos; parameters are indicated by white circles in (b).    
    Negative echos decrease the radius of gyration by introducing a crossover regime where the curves plateau (black arrow).
    }
    \label{fig::time-delayed_wormlike-chain}
\end{figure}

In light of these results, we now revisit the scenario of a wormlike chain subject to temporal sequences of kicks, but this time assume that each excitation has a characteristic size (correlation length) comparable to the length of a Kuhn segment, $l_e \sim 2l_p$.
In practice, we approximate this by a discrete semiflexible polymer~\cite{supplement, Article::Winkler2003, Article::Tejedor2022} where each coarse-grained monomer corresponds to a Kuhn length, so that the excitations at different monomers can be assumed independent.
As shown in Fig.~\ref{fig::time-delayed_wormlike-chain}(a), effective bending of active regions~\cite{Article::Goychuk2023}, combined with the constraint of chain inextensibility, drastically reduces the mean squared separation between the boundaries of the active region, which is indicative of a loop-like state.
Compared to a scenario without echos, positive echos deplete contacts within this loop-like domain [above the diagonal in Fig.~\ref{fig::time-delayed_wormlike-chain}(a)].
Conversely, negative echos enhance contacts and can even induce compaction by almost $100\%$, reminiscent of a coil-to-globule transition [below the diagonal in Fig.~\ref{fig::time-delayed_wormlike-chain}(a)].
Negative echos can be viewed as time-delayed analogs of local force dipoles, where the delay originates from stress propagation through internal degrees of freedom.
For large molecular motors such as cohesin, for example, the force exerted by one motor head as it extrudes a chromatin loop~\cite{Review::Banigan}, could propagates along the clamped chromatin loop or the body of the cohesin molecule, and would be echoed by an equal and opposite force at the same locus.
Moreover, we expect that time-delayed echos can also arise from interaction forces during local sequences of chemical reactions.

In summary, we have analyzed the conformations and dynamics of a broad class of inhomogeneous active polymers which experience local temporal patterns of excitations.
Within this theoretical framework, different segments can vary in activity, persistence time, or even show unique temporal excitation programs.
Then, both qualitative features of the excitations (such as, positive or negative echos) and quantitative features (delay times of the echos) determine which regions of the polymer move coherently and how the active segment folds.
Moreover, our framework can explain previous theoretical observations of polymer swelling and collapse~\cite{Article::Eisenstecken2016, Article::Eisenstecken2017a, Article::Mousavi2019, Article::Anand2020, Article::Brahmachari2023, Review::Winkler2020} as a competition between a) the polymer persistence length, and b) the typical arc distance that excitations are relayed through tension propagation along the polymer backbone.
We hypothesize that such a competition of length scales could be relevant for biopolymers such as chromatin, and could determine if active segments expand or collapse.


\begin{acknowledgements}
We thank Arup K. Chakraborty for engaging in insightful discussions.
This work was supported by the National Science Foundation, through the Biophysics of Nuclear Condensates grant (MCB-2044895).
A.G. was also supported by an EMBO Postdoctoral Fellowship (ALTF 259-2022).
D.K. was supported by the Graduate Research Fellowship Program under grant No. 2141064.
The authors acknowledge the MIT SuperCloud and Lincoln Laboratory Supercomputing Center for providing HPC resources that have contributed to the research results reported within this paper.
\end{acknowledgements}

\bibliographystyle{apsrev4-2}
\bibliography{bibliography}

\end{document}


\title{Delayed excitations induce polymer looping and coherent motion}

\author{Andriy Goychuk}
\email{andriy@goychuk.me}
\affiliation{Institute for Medical Engineering and Science, Massachusetts Institute of Technology, Cambridge, MA 02139, United States}%
\author{Deepti Kannan}
\affiliation{Department of Physics, Massachusetts Institute of Technology, Cambridge, MA 02139, United States}%
\author{Mehran Kardar}
\email{kardar@mit.edu}
\affiliation{Department of Physics, Massachusetts Institute of Technology, Cambridge, MA 02139, United States}%

\maketitle

\clearpage
\tableofcontents
\clearpage

\section{Time-dependent correlation between Rouse modes}
\label{sec::correlation_full}
In the present section, we determine the matrix of correlations between time-lagged Rouse modes.
As discussed in the main text, the linear response of the Rouse matrix $\mat{R}(t)$ to random excitations $\mat{H}(t)$ with covariance $\mat{C}(t-t') \coloneqq \langle\mat{H}(t)\cdot\mat{H}^\dagger(t')\rangle$ is characterized by
\begin{equation}
    \mat{R}(t) = \int_{-\infty}^{t} \! d\tau \, e^{-\mat{J} (t-\tau)} \cdot \mat{H}(\tau) \, ,
    \label{si-eq::solution_formal}
\end{equation}
where $\mat{J}$ is the response matrix.
Note that Eq.~\eqref{si-eq::solution_formal} and the subsequent discussion can be interpreted in (i) Fourier space or in (ii) real space.
In the first case, $\mat{R}(t)$ and $\mat{H}(t)$ are the matrices of Rouse modes and excitation modes, respectively, while $\mat{J}$ represents the mechanical coupling between different modes.
In the second case, the matrices $\mat{R}(t)$ and $\mat{H}(t)$ contain the position and excitation vectors of all monomers, while $\mat{J}$ represents the mechanical coupling between different monomers.
In our analytic calculations, such as for continuous chains discussed in Sec.~\ref{sec::inextensibility_weak}, we use (i) the Rouse mode decomposition.
In our numerical evaluations, such as for discrete chains discussed in Sec.~\ref{sec:discrete_wormlike_chain}, we use (ii) a real-space representation.

Now, we will show how the correlations between time-lagged Rouse matrices, $\mat{X}(t-t') \coloneqq \langle \mat{R}(t) \cdot \mat{R}^\dagger(t')\rangle$, can be expressed as the solution a Lyapunov equation.
To that end, we first multiply Eq.~\eqref{si-eq::solution_formal} with its conjugate transpose, and average over the noise:
\begin{equation}
    \langle\mat{R}(t) \cdot \mat{R}^\dagger(t')\rangle = 
    \int_{-\infty}^{t} \! d\tau \, 
    \int_{-\infty}^{t'} \! d\tau' \,
    e^{-\mat{J} (t-\tau)} \cdot 
    \langle\mat{H}(\tau) \cdot \mat{H}^\dagger(\tau') \rangle \cdot 
    e^{-\mat{J}^\dagger (t'-\tau')}
    \, .
    \label{si-eq:correlations_step_1}
\end{equation}
Next, we substitute the covariance of the noise, $\mat{C}(\tau-\tau') \coloneqq \langle\mat{H}(\tau)\cdot\mat{H}^\dagger(\tau')\rangle$ and integrate by parts (along $d\tau$):
\begin{multline}
    \langle\mat{R}(t) \cdot \mat{R}^\dagger(t')\rangle = 
    \mat{J}^{-1} 
    \int_{-\infty}^{t'} \! d\tau' \, 
    \mat{C}(t-\tau') \cdot 
    e^{-\mat{J}^\dagger (t'-\tau')} \\
    -\mat{J}^{-1}
    \int_{-\infty}^{t} \! d\tau \, 
    \int_{-\infty}^{t'} \! d\tau' \,
    e^{-\mat{J} (t-\tau)} \cdot 
    \partial_\tau\mat{C}(\tau-\tau') \cdot 
    e^{-\mat{J}^\dagger (t'-\tau')}
    \, .
\end{multline}
Using $\partial_\tau\mat{C}(\tau-\tau') = -\partial_{\tau'} \mat{C}(\tau-\tau')$, and again integrating by parts (this time along $d\tau'$) gives
\begin{multline}
    \langle\mat{R}(t) \cdot \mat{R}^\dagger(t')\rangle = 
    \mat{J}^{-1} 
    \int_{-\infty}^{t'} \! d\tau' \,
    \mat{C}(t-\tau') \cdot 
    e^{-\mat{J}^\dagger (t'-\tau')}
    +\mat{J}^{-1}
    \int_{-\infty}^{t} \! d\tau \, 
    e^{-\mat{J} (t-\tau)} \cdot 
    \mat{C}(\tau-t') \\
    -\mat{J}^{-1}
    \int_{-\infty}^{t} \! d\tau \, 
    \int_{-\infty}^{t'} \! d\tau' \,
    e^{-\mat{J} (t-\tau)} \cdot 
    \mat{C}(\tau-\tau') \cdot 
    e^{-\mat{J}^\dagger (t'-\tau')} \cdot \mat{J}^\dagger
    \, .
\end{multline}
Substitution of Eq.~\eqref{si-eq:correlations_step_1} into the second line, substitution of $\mat{X}(t-t') \coloneqq \langle \mat{R}(t) \cdot \mat{R}^\dagger(t')\rangle$, and rearrangement of the terms leads to:
\begin{multline}
    \mat{J}\cdot\mat{X}(t-t') + \mat{X}(t-t') \cdot \mat{J}^\dagger = 
    \int_{-\infty}^{t'} \! d\tau' \,
    \mat{C}(t-\tau') \cdot 
    e^{-\mat{J}^\dagger (t'-\tau')}
    + \int_{-\infty}^{t} \! d\tau \, 
    e^{-\mat{J} (t-\tau)} \cdot 
    \mat{C}(\tau-t') \, .
\end{multline}
Finally, after a change of variables in both integrals, and defining $\Delta t \coloneqq t-t'$, one arrives at the following Lyapunov equation:
\begin{equation}    
    \mat{J}\cdot \mat{X}(\Delta t) + \mat{X}(\Delta t) \cdot \mat{J}^\dagger 
    = \int_{\Delta t}^{\infty} \!\!\! d\tau \, \mat{C}(\tau) \cdot e^{-\mat{J}^\dagger |\tau - \Delta t|} 
    + \int_{-\infty}^{\Delta t} \!\!\! d\tau \, e^{-\mat{J} |\tau - \Delta t|} \cdot \mat{C}(\tau)
    \, .
    \label{si-eq::lyapunov_effective}
\end{equation}
%
The steady state of the polymer, which corresponds to $\Delta t = 0$, is then characterized by
\begin{equation}    
    \mat{J}\cdot \mat{X}(0) + \mat{X}(0) \cdot \mat{J}^\dagger 
    = \int_{0}^{\infty} \!\!\! d\tau \, \left[ \mat{C}(\tau) \cdot e^{-\mat{J}^\dagger \tau} 
    + e^{-\mat{J} \tau} \cdot \mat{C}^\dagger(\tau) \right] 
    \coloneqq
    \mat{C}_\textit{eff}
    \, .
    \label{si-eq::lyapunov_effective_steady_state}
\end{equation}
Next, we analyze the fluctuations for large time lags $\Delta t$.
Because $\mat{X}(-\Delta t) = \mat{X}^\dagger(\Delta t)$, the subsequent discussion focuses on the case $\Delta t > 0$ without loss of generality.

We consider a scenario where the covariance matrix of the excitations, $\mat{C}(\tau)$, has a characteristic decay time $\tau_c$ so that $\mat{C}(\tau) \sim 0$ for all $|\tau| > \tau_c$. 
This implies that the integrands in Eqs.~\eqref{si-eq::lyapunov_effective}~and~\eqref{si-eq::lyapunov_effective_steady_state}, as well as in the remainder of this section, vanish for $|\tau| > \tau_c$.
In this context, time integration bounds that lie at $\pm\infty$ can be freely interchanged with integration bounds that lie at $\pm\tau_c$.
To further simplify our description, we focus on large time lags $|\Delta t| \gg \tau_c$, so that the polymer has ample time to run through many excitation realizations.
With this simplification, one has
\begin{equation}    
    \mat{J}\cdot \mat{X}(\Delta t) + \mat{X}(\Delta t) \cdot \mat{J}^\dagger 
    = e^{-\mat{J} \, |\Delta t|} \cdot \int_0^{\tau_c} \!\!\! d\tau \, \left[ e^{-\mat{J}\tau}\cdot\mat{C}^\dagger(\tau) + e^{\mat{J}\tau}\cdot \mat{C}(\tau) \right] \, .
    \label{si-eq::lyapunov_fluctuation_1}
\end{equation}
Next, we slightly rewrite Eq.~\eqref{si-eq::lyapunov_fluctuation_1} to
\begin{equation}    
    \mat{J}\cdot \mat{X}(\Delta t) + \mat{X}(\Delta t) \cdot \mat{J}^\dagger 
    = e^{-\mat{J} \, |\Delta t|} \cdot \left\{ \mat{C}_\textit{eff} +
    \int_0^{\tau_c} \!\!\! d\tau \, \left[e^{\mat{J}\tau}\cdot \mat{C}(\tau) - \mat{C}(\tau) \cdot e^{-\mat{J}^\dagger\tau} \right] \right\} \, .
    \label{si-eq::lyapunov_fluctuation_2}
\end{equation}
Finally, we remark that the magnitude of the response matrix is inversely proportional to the microscopic timescale, $\mat{J}\propto\tau_b^{-1}$.
Using this proportionality and expanding the matrix exponentials in Eq.~\eqref{si-eq::lyapunov_fluctuation_2} reveals that 
\begin{equation}    
    \mat{J}\cdot \mat{X}(\Delta t) + \mat{X}(\Delta t) \cdot \mat{J}^\dagger 
    = e^{-\mat{J} \, |\Delta t|} \cdot \Bigl[ \mat{C}_\textit{eff} +
    \mathcal{O}(\tau_c/\tau_b) \Bigr] \, .
    \label{si-eq::lyapunov_fluctuation_3}
\end{equation}
Thus, for $\tau_c \ll \tau_b$, the fluctuations of the chain can be approximated by
\begin{equation}    
    \mat{X}(\Delta t)  
    \approx 
    \begin{cases}
    e^{-\mat{J} \, |\Delta t|} \cdot \mat{X}(0) \, , & \text{for} \quad \Delta t \geq 0 \, , \\
    \mat{X}(0) \cdot e^{-\mat{J}^\dagger \, |\Delta t|} \, , & \text{for} \quad \Delta t < 0 \, .
    \end{cases}
    \label{si-eq::lyapunov_fluctuation_4}
\end{equation}
In all other scenarios, where the above simplifications cannot be made, we evaluate Eq.~\eqref{si-eq::lyapunov_effective} numerically.

\vfill
\newpage

\section{Continuous wormlike chains}
\label{sec::inextensibility_weak}

\subsection{Enforcing the constraint of weak inextensibility}

In the present section, we derive an approximation for the Lagrange multiplier $\kappa(s)$, which can be interpreted as an inhomogeneous line tension that weakly enforces the inextensibility of active wormlike chains on the ensemble-averaged level.
We refer to this constraint as ``weak'' because it can be broken on the level of individual polymer trajectories.

\subsubsection{Excitations with finite correlation length}

Analogous to the main text, for simplicity we consider a scenario where the modulations in $\mathcal{G}(s,t)$ have a much larger length scale than the correlation length of the excitations, $\partial_s\mathcal{G} \ll l_e^{-1} \mathcal{G}$.
Then, the covariance of the excitations can be well approximated by 
\begin{equation}
    \langle\vec{\eta}(s, t) \cdot \vec{\eta}(s', t')\rangle = \frac{2b^2}{\tau_b} 
    \frac{1}{2l_e} \, e^{-\frac{|s-s'|}{l_e}} \, 
    \widehat{\mathcal{C}}\bigl(\tfrac{s+s'}{2}, t-t'\bigr) \, .
\end{equation}
To simplify our description in the following, we take the limit of a very long chain.
In Fourier space, the pairwise correlation between excitation modes is therefore given by
\begin{equation}
    \langle\vec{\eta}_q(t)\cdot \vec{\eta}^\ast_k(t')\rangle = \frac{2b^2}{\tau_b} \, \frac{4}{4+l_e^2 (q+k)^2} \, \widehat{C}_{q-k}(t-t') \, ,
\end{equation}
where $\widehat{C}_{q}(\Delta t)$ is the Fourier transform of $\widehat{\mathcal{C}}(s, \Delta t)$.
To further simplify our discussion, we focus on a scenario where there are no time-delayed excitation echos, so that 
\begin{equation}
    \widehat{C}_{q}(\Delta t) = \bigl[ L \delta_{q,0} + \epsilon_{q} \bigr] \, \delta(\Delta t) \, ,
\end{equation}
where $L$ is the (dimensionless) polymer length.
Thus, the full expression for the correlation between Rouse modes is $\langle \mat{H}(t) \cdot \mat{H}^\dagger(t')\rangle = \mat{C}^{(0)} \, \delta(t-t')$ with the covariance matrix
\begin{equation}
    C^{(0)}_{qk} = \frac{2b^2}{\tau_b} \, \frac{4}{4+l_e^2 (q+k)^2} \, \bigl[ L \delta_{qk} + \epsilon_{q-k} \bigr] \, .
    \label{si-eq::excitation_correlation_simplified}
\end{equation}
Given these excitations, we will next determine the resulting steady state of the wormlike chain.

\subsubsection{Steady state of the wormlike chain}
As discussed in the main text, the response matrix of the active wormlike chain under consideration is given by $\tau_b J_{qk} = \frac{1}{L} q k \, \kappa_{q-k} + \tfrac{1}{4} q^4 \delta_{qk}$.
Since the polymer is subject to excitations without memory, Eq.~\eqref{si-eq::excitation_correlation_simplified}, the correlation between Rouse modes in steady state is given by the following simplified expression~\cite{Article::Goychuk2023}:
\begin{equation}
    \langle \mat{R}(t) \cdot \mat{R}^\dagger(t) \rangle = \int_{0}^{\infty} \! d\tau \, e^{-\mat{J}\tau}\cdot\mat{C}^{(0)}\cdot e^{-\mat{J}^\dagger\tau}\, , 
\end{equation}
which follows from Eq.~\eqref{si-eq::solution_formal}.
To calculate this expression, we use an expansion for the matrix exponential, which treats the off-diagonal elements introduced by the Lagrange multiplier $\epsilon_{q-k}$ as a small perturbation.
In the following, we set $\epsilon_0 = 0$ and assume that $L\gg l_e$.
Using this perturbative approach, as outlined in our prior work~\cite{Article::Goychuk2023}, one has for the tangent vectors $\vec{\tau}(s,t) \coloneqq b^{-1} \partial_s \vec{r}(s,t)$:
\begin{equation}
    \langle \vec{\tau}_q(t) \cdot \vec{\tau}_k^\ast(t)\rangle = 
    \frac{qk}{1+l_e^2 q^2} \frac{L\delta_{qk}}{f(q)}
    + \frac{8 \, qk}{4+l_e^2(q+k)^2} \frac{\epsilon_{q-k}}{f(q)+f(k)} 
    - (qk)^2 \, \frac{\kappa_{q-k}-\kappa_0\delta_{qk}}{f(q)f(k)} \, ,
    \label{si-eq::tangent_covariance}
\end{equation}
where we have defined $f(q) \coloneqq \frac{\kappa_0}{L} q^2 + \frac{1}{4} q^4$.
We posit that the tangent vectors are normalized on average, $\langle |\vec{\tau}(s,t)|^2 \rangle = 1$, which implies
\begin{equation}
    \frac{1}{L}\sum_q \langle \vec{\tau}_q(t) \cdot \vec{\tau}^\ast_{q-k}(t)\rangle = L \delta_{k,0} \, .
    \label{si-eq::inextensibility_constraint}
\end{equation}
Comparing Eq.~\eqref{si-eq::tangent_covariance} and Eq.~\eqref{si-eq::inextensibility_constraint}, one finds the following set of conditions that \emph{all} need to be fulfilled to enforce inextensibility:
\begin{subequations}
\begin{align}
    \label{si-eq::condition_1}
    k = 0 : \quad & \int_{-\infty}^{\infty}\!\frac{dq}{2\pi}\,\frac{q^2}{1+l_e^2 q^2} \frac{1}{f(q)} = 1 \, , \\
    \label{si-eq::condition_2}
    k \neq 0 : \quad & \left[ \int_{-\infty}^{\infty} \!dq\, \frac{q^2(q-k)^2}{f(q)f(q-k)}\right] \kappa_k =
    \left[\int_{-\infty}^{\infty} \!dq\, \frac{8}{4+l_e^2(2q-k)^2} \frac{q(q-k)}{f(q) + f(q-k)} \right] \epsilon_k \, ,
\end{align}
\end{subequations}
where we have approximated the sums via integrals.
From the first condition, Eq.~\eqref{si-eq::condition_1}, it follows that 
\begin{equation}
    \chi(l_e) \coloneqq \left(\frac{\kappa_0}{L}\right)^{-1} = \frac{1 + 4 l_e + \sqrt{1 + 8 l_e}}{2} \, .
\end{equation}
The second set of conditions, Eq.~\eqref{si-eq::condition_2}, relates the modulations of the Lagrange multiplier $\kappa(s)$ to the activity modulations $\epsilon(s)$.
For large wave modes, $k \rightarrow 0$ which corresponds to $\partial_s \epsilon(s) \ll l_e^{-1}\epsilon(s)$, one has
\begin{equation}
    \kappa(s) = \frac{1}{\chi(l_e)} + \frac{2\epsilon(s)}{\chi(l_e)^{3/2}} - \frac{\partial_s^2 \epsilon(s)}{8 \, \chi(l_e)^{1/2}} \, ,
\end{equation}
up to the leading terms in the Lagrange multiplier $\kappa(s)$.
Note that the last term is sub-leading when activity modulations occur over sequences larger than $l_e$.

\vfill
\newpage

\section{Discrete wormlike chains}
\label{sec:discrete_wormlike_chain}

\subsection{Response matrix}
In this section, following Refs.~\cite{Article::Winkler2003, Article::Tejedor2022}, we describe the response matrix of a discrete wormlike chain consisting of $N$ monomers.
We assume that the sequence space distance between adjacent monomers is given by $ds$, thus defining the discretization of the polymer.
In other words, each coarse-grained monomer contains $ds$ Kuhn lengths of a microscopic chain.
The tangent vectors, $\vec{\tau}_i \coloneqq b^{-1} \, ds^{-1}[\vec{r}_{i+1}-\vec{r}_i]$, must satisfy the following constraints:
\begin{subequations}    
\begin{align}
    \label{eq:discrete_inextensibility}
    \langle |\vec{\tau}_i|^2 \rangle & \equiv 1 \, , \\
    \langle \vec{\tau}_i \cdot \vec{\tau}_{i+1} \rangle &\equiv e^{-\frac{ds}{l_p}} = e^{-2 \, ds} \, ,
\end{align}
\end{subequations}
which encode chain inextensibility and the distribution of bond angles at each monomer.
For a compact notation, we first introduce the discrete derivative operator as a ${N{-}1 \times N}$ matrix whose elements are given by
\begin{equation}
    \left(\discretenabla\right)_{ij} \coloneqq
    ds^{-1} \bigl[\delta_{i+1,j} - \delta_{ij}\bigr] \, .
\end{equation}
The ${N{-}1 \times 3}$ matrix of all tangent vectors, $\mat{T}$ with rows $T_{i,\dots} \coloneqq \vec{\tau}_i$, is then given by $\mat{T} = b^{-1} \, \discretenabla \cdot \vec{R}$, where the matrix of all position vectors $\mat{R}$ has rows $R_{i,\dots} \coloneqq \vec{r}_i$.

Next, we define the diagonal ${N{-}1 \times N{-}1}$ stiffness matrix whose elements,
\begin{equation}
    K_{ij} \coloneqq \kappa_i \, \delta_{ij} \, ,
\end{equation}
indicate the stiffness $\kappa_i$ of the spring connecting beads $i$ and $i{+}1$, and the off-diagonal ${N{-}1 \times N{-}1}$ bending matrix whose elements,
\begin{equation}
    B_{ij} \coloneqq \frac{1}{2} \, \bigl[ \omega_i \, \delta_{i+1,j} + \omega_j \, \delta_{i,j+1} \bigr] \, ,
\end{equation}
indicate the angular stiffness $\omega_i$ of the bonds at bead $i{+}1$.
The response matrix can be derived starting from a chain configuration energy as described in Ref.~\cite{Article::Winkler2003, Article::Tejedor2022} and is given by
\begin{equation}
    \tau_b \, \mat{J} = \discretenabla^T \cdot \bigl[\mat{K} - \mat{B}\bigr] \cdot \discretenabla \, .
\end{equation}
As shown in Refs.~\cite{Article::Winkler2003, Article::Tejedor2022}, the stiffness parameters for a homogeneous wormlike chain which satisfies Eq.~\eqref{eq:discrete_inextensibility} are given by
\begin{subequations}
\label{eq:initial_guess}
\begin{equation}
    \kappa_i = \frac{1}{2} \frac{1}{1-e^{-4 \,ds}} \color{gray}\times\color{black} \begin{cases}
        1 \, , & i \in \{1,N-1\} \, , \\
        1 + e^{-4 \, ds} \, , & \text{else} \, ,
    \end{cases}
\end{equation}
while the bond angle stiffness is homogeneous,
\begin{equation}
    \omega_i = \frac{e^{-2 \, ds}}{1- e^{-4 \, ds}} \, .
\end{equation}
\end{subequations}
Based on this description, Ref.~\cite{Article::Tejedor2022} has studied thermal chains with position-dependent stiffness.
Here, we here focus on active polymers.
Starting from Eq.~\eqref{eq:initial_guess} as initial guess and keeping the bending rigidity $\omega_i$ constant, we then iteratively optimize the spring stiffnesses $\kappa_i$ to satisfy the inextensibility constraint Eq.~\eqref{eq:discrete_inextensibility}.

\subsection{Excitations effectively have a finite correlation length}

At every bead, we apply statistically independent excitations.
Nevertheless, in contrast to the continuum model, the coarse-grained nature of the discrete chain suggests that each bead could accommodate multiple persistence lengths of a finer-grained microscopic polymer.
Then, each point of the finer-grained microscopic polymer that lies within a bead of the coarse-grained Rouse chain will experience correlated excitations.
This suggests that the excitations have an effective correlation length $l_e \sim ds$.
While this heuristic comparison is not meant to be exact, in part because the excitations in the continuum model are not box- but exponentially correlated in sequence space, it provides a qualitative indicator of what $l_e \gg l_p$ or $l_e \ll l_p$ means for the discrete chain: $ds \gg 1$ or $ds \ll 1$, respectively.

\vfill
\newpage

\section{Parameters}

This section serves as a central reference that provides all model parameters, response functions, and memory kernels for the excitations, which were used in generating the figures in this work.
All parameters are given in units of the Kuhn length $b$ and the microscopic relaxation time $\tau_b$.
As discussed in the main text, the covariance of the excitations in the continuum models are given by
\begin{subequations}
\begin{equation}
    \langle\vec{\eta}(s, t) \cdot \vec{\eta}(s', t')\rangle = \frac{2b^2}{\tau_b} 
    \frac{1}{2l_e} \, e^{-\frac{|s-s'|}{l_e}} 
    \widehat{\mathcal{C}}\bigl(\tfrac{s+s'}{2}, t-t'\bigr) \, .
\end{equation}
For our discrete models, we use the analogous form
\begin{align}
    \langle\vec{\eta}_n(t) \cdot \vec{\eta}_m(t')\rangle &= \frac{2b^2}{\tau_b} \delta_{nm} \, \widehat{\mathcal{C}}\bigl(n, t-t'\bigr) \, .
\end{align}
Thus, discrete monomers experience pairwise independent excitations.
However, as discussed in Sec.~\ref{sec:discrete_wormlike_chain} each coarse-grained monomer contains multiple Kuhn lengths of a finer-grained microscopic polymer.
Thus, the chain discretization $ds$ sets the effective correlation length of the excitations.
The temporal correlations are, as discussed in the main text, given by
\begin{equation}
    \widehat{\mathcal{C}}(s,t) = \delta(t) + \lim_{\varepsilon \to 0}  \int_{-\varepsilon}^{\infty} \! d\tau \, \mathcal{G}(s, |t| + \tau) \, \mathcal{G}(s, \tau) \, .
\end{equation}
\end{subequations}
where $\mathcal{G}(s,t)$ is the memory kernel of the excitations.

\vfill
\newpage

\begin{table}[h]
\caption{%
Parameters used for \textbf{Figure 2}.\hfill
}
\begin{ruledtabular}
\begin{tabular}{p{0.35\linewidth}p{0.65\linewidth}}
\textbf{model type} & discrete flexible (Rouse) chain \\
\hline
\textbf{chain discretization} & 
$ds = 1$ \\
\textbf{chain dimensions} & 
$s \equiv ds \, n$ \newline 
$n \in [1,1000]$ \\
\textbf{plot range} & 
$n \in [300, 700]$ \\
\textbf{equation of motion} & 
\(\displaystyle
\partial_t\vec{r}_n(t) - \vec{\eta}_n(t) = - \sum_m J_{nm} \vec{r}_m(t)
\) \\
\textbf{response matrix} \newline (real space) & 
\(\displaystyle
\tau_b \mat{J} = -\tfrac{1}{2}\discretenabla^T \cdot \discretenabla
\) \\
\hline
\textbf{excitation properties} \\
\hline
\textbf{memory kernel} & 
\(\displaystyle
\mathcal{G}(n,t) = \sqrt{\epsilon(n)} \, [\delta(t) + \epsilon_c \, \delta(t - \tau_c)]
\) \\
\textbf{activity profile} &
\(\displaystyle
\epsilon(n) =
1.0 \color{gray}\times\color{black} \begin{cases}
1 \, , & n \in [451, 551] \, ,  \\
0 \, , & \text{else}
\end{cases}
\) \\
\textbf{echo strength} & 
above diagonal: $\epsilon_c = +1$ \newline
below diagonal: $\epsilon_c = -1$ \\
\textbf{delay time} & 
panel 1: $\tau_c = 10^1 \, \tau_b$ \newline
panel 2: $\tau_c = 10^2 \, \tau_b$ \newline
panel 3: $\tau_c = 10^3 \, \tau_b$ \newline
panel 4: $\tau_c \to \infty$ \\
\end{tabular}
\end{ruledtabular}
\end{table}

\vfill
\newpage

\begin{table}[h]
\caption{%
Parameters used for \textbf{Figure 3b)}, \textbf{above the diagonal}.\hfill
}
\begin{ruledtabular}
\begin{tabular}{p{0.35\linewidth}p{0.65\linewidth}}
\textbf{model type} & 
continuous semiflexible (wormlike) chain \newline 
large length scales where tangent vector alignment due to bending rigidity can be neglected. \\
\hline
\textbf{chain dimensions} & 
$\displaystyle s \in [-L/2, L/2]$ \newline
$L\to\infty$ \\
\textbf{plot range} & 
$\displaystyle s \in [-1, 1]$ \\
\textbf{equation of motion} & 
\(\displaystyle
\tau_b\bigl[\partial_t\vec{r}(s,t) - \vec{\eta}(s,t) \bigr] = \partial_s \! \left[ \left(\frac{1}{\chi(l_e)} + \frac{2\epsilon(s)}{\chi(l_e)^{3/2}} \right) \partial_s\vec{r}(s,t) \right]
\) \\
\textbf{response matrix} \newline (Fourier space) & 
$\displaystyle \tau_b \, J_{qk} = \frac{1}{\chi(l_e)} q^2 \delta_{qk} + \frac{2}{\chi(l_e)^{3/2}} \frac{\epsilon_{q-k}}{L} \, q k $ \\
\hline
\textbf{excitation properties} \\
\hline
\textbf{correlation length} & 
panel 1: $l_e = 0$ \newline
panel 2: $l_e = 2 l_p$ \newline
panel 3: $l_e = 20 l_p$ \\
\textbf{memory kernel} & 
$\mathcal{G}(s,t) = \sqrt{\epsilon(s)} \, \delta(t)$ \\
\textbf{activity profile} &
$\epsilon(s) =
0.5 \color{gray}\times\color{black} \begin{cases}
1 \, , & s \in [-0.4, 0.4] \, ,  \\
0 \, , & \text{else}
\end{cases}$ \\
\end{tabular}
\end{ruledtabular}
\end{table}

\vfill
\newpage

\begin{table}[h]
\caption{%
Parameters used for \textbf{Figure 3b)}, \textbf{below the diagonal}.\hfill
}
\begin{ruledtabular}
\begin{tabular}{p{0.35\linewidth}p{0.65\linewidth}}
\textbf{model type} & discrete semiflexible (wormlike) chain \\
\hline
\textbf{chain discretization} & 
panel 1: $ds = 1 \color{gray}\times\color{black} 10^{-1}$ \newline
panel 2: $ds = 1 \color{gray}\times\color{black} 10^{0}$ \newline
panel 3: $ds = 1 \color{gray}\times\color{black} 10^{3} $ \\
\textbf{chain dimensions} & 
$s \equiv ds \, n$ \newline
$n \in [1,1000]$ \\
\textbf{plot range} & 
$n \in [1, 1000]$ \\
\textbf{equation of motion} & 
\(\displaystyle
\partial_t\vec{r}_n(t) - \vec{\eta}_n(t) = - \sum_m J_{nm} \vec{r}_m(t)
\) \\
\textbf{response matrix} \newline 
(real space) & 
\(\displaystyle
\tau_b \, \mat{J} = \discretenabla^T \cdot \bigl[\mat{K} - \mat{B}\bigr] \cdot \discretenabla
\) \\
\hline
\textbf{excitation properties} \\
\hline
\textbf{correlation length} & effectively set by chain discretization \\
\textbf{memory kernel} & 
\(\displaystyle
\mathcal{G}(n,t) = \sqrt{\epsilon(n)} \, \delta(t)
\) \\
\textbf{activity profile} &
\(\displaystyle
\epsilon(n) =
0.5 \color{gray}\times\color{black}\begin{cases}
1 \, , & n \in [301, 701] \, ,  \\
0 \, , & \text{else}
\end{cases}
\) \\
\end{tabular}
\end{ruledtabular}
\end{table}

\vfill
\newpage

\begin{table}[h]
\caption{%
Parameters used for \textbf{Figure 4a)}.\hfill
}
\begin{ruledtabular}
\begin{tabular}{p{0.35\linewidth}p{0.65\linewidth}}
\textbf{model type} & discrete semiflexible (wormlike) chain \\
\hline
\textbf{chain discretization} & 
$ds = 1$ \\
\textbf{chain dimensions} & 
$s \equiv ds \, n$ \newline 
$n \in [1,1000]$ \\
\textbf{plot range} & 
$n \in [1, 1000]$ \\
\textbf{equation of motion} & 
\(\displaystyle
\partial_t\vec{r}_n(t) - \vec{\eta}_n(t) = - \sum_m J_{nm} \vec{r}_m(t)
\) \\
\textbf{response matrix} & 
\(\displaystyle
\tau_b \, \mat{J} = \discretenabla^T \cdot \bigl[\mat{K} - \mat{B}\bigr] \cdot \discretenabla
\) \\
\hline
\textbf{excitation properties} \\
\hline
\textbf{correlation length} & effectively set by chain discretization \\
\textbf{memory kernel} & 
\(\displaystyle
\mathcal{G}(n,t) = \sqrt{\epsilon(n)} \, [\delta(t) + \epsilon_c \, \delta(t - \tau_c)]
\) \\
\textbf{activity profile} &
\(\displaystyle
\epsilon(n) =
1.0 \color{gray}\times\color{black} \begin{cases}
1 \, , & n \in [301, 701] \, ,  \\
0 \, , & \text{else}
\end{cases}
\) \\
\textbf{echo strength} & 
above diagonal: $\epsilon_c = +1$ \newline
below diagonal: $\epsilon_c = -1$ \\
\textbf{delay time} & 
panel 1: $\tau_c = 10^0 \, \tau_b$ \newline
panel 2: $\tau_c = 10^2 \, \tau_b$ \newline
panel 3: $\tau_c = 10^4 \, \tau_b$ \newline
panel 4: $\tau_c \to \infty$ \\
\end{tabular}
\end{ruledtabular}
\end{table}

\vfill
\newpage

\begin{table}[h]
\caption{%
Parameters used for \textbf{Figure 4b)}.\hfill
}
\begin{ruledtabular}
\begin{tabular}{p{0.35\linewidth}p{0.65\linewidth}}
\textbf{model type} & discrete semiflexible (wormlike) chain \\
\hline
\textbf{chain discretization} & 
$ds = 1$ \\
\textbf{chain dimensions} & 
$s \equiv ds \, n$ \newline 
$n \in [1,1000]$ \\
\textbf{equation of motion} & 
\(\displaystyle
\partial_t\vec{r}_n(t) - \vec{\eta}_n(t) = - \sum_m J_{nm} \vec{r}_m(t)
\) \\
\textbf{response matrix} & 
\(\displaystyle
\tau_b \, \mat{J} = \discretenabla^T \cdot \bigl[\mat{K} - \mat{B}\bigr] \cdot \discretenabla
\) \\
\hline
\textbf{excitation properties} \\
\hline
\textbf{correlation length} & effectively set by chain discretization \\
\textbf{memory kernel} & 
\(\displaystyle
\mathcal{G}(n,t) = \sqrt{\epsilon(n)} \, [\delta(t) + \epsilon_c \, \delta(t - \tau_c)]
\) \\
\textbf{activity profile} &
\(\displaystyle
\epsilon(n) =
E \color{gray}\times\color{black} \begin{cases}
1 \, , & n \in [301, 701] \, ,  \\
0 \, , & \text{else}
\end{cases}
\) \\
\textbf{maximum activity} &
$E\in [0.01,20]$ \\
\textbf{echo strength} & 
left panel: $\epsilon_c = +1$ \newline
right panel: $\epsilon_c = -1$ \\
\textbf{delay time} & 
$\tau_c \in [10^{-4}, 8.9 \times 10^{7}] \color{gray}\times\color{black} \tau_b$ \\
\end{tabular}
\end{ruledtabular}
\end{table}

\vfill
\newpage

\begin{table}[h]
\caption{%
Parameters used for \textbf{Figure 4c)}.\hfill
}
\begin{ruledtabular}
\begin{tabular}{p{0.35\linewidth}p{0.65\linewidth}}
\textbf{model type} & discrete semiflexible (wormlike) chain \\
\hline
\textbf{chain discretization} & 
$ds = 1$ \\
\textbf{chain dimensions} & 
$s \equiv ds \, n$ \newline 
$n \in [1,1000]$ \\
\textbf{equation of motion} & 
\(\displaystyle
\partial_t\vec{r}_n(t) - \vec{\eta}_n(t) = - \sum_m J_{nm} \vec{r}_m(t)
\) \\
\textbf{response matrix} & 
\(\displaystyle
\tau_b \, \mat{J} = \discretenabla^T \cdot \bigl[\mat{K} - \mat{B}\bigr] \cdot \discretenabla
\) \\
\hline
\textbf{excitation properties} \\
\hline
\textbf{correlation length} & effectively set by chain discretization \\
\textbf{memory kernel} & 
\(\displaystyle
\mathcal{G}(n,t) = \sqrt{\epsilon(n)} \, [\delta(t) + \epsilon_c \, \delta(t - \tau_c)]
\) \\
\textbf{activity profile} &
\(\displaystyle
\epsilon(n) =
E \color{gray}\times\color{black} \begin{cases}
1 \, , & n \in [1, 1000] \, ,  \\
0 \, , & \text{else}
\end{cases}
\) \\
\textbf{maximum activity} &
$E\in \{0.25, 0.5, 1.0, 2.0, 4.0, 8.0\}$ \\
\textbf{echo strength} & 
$\epsilon_c = -1$ \\
\textbf{delay time} & 
$\tau_c = 10^{0} \, \tau_b$ \\
\end{tabular}
\end{ruledtabular}
\end{table}

\vfill
\newpage

\bibliography{bibliography}